\documentclass[prd,aps,preprint,tightenlines,nofootinbib,superscriptaddress]{revtex4}
\usepackage{mathrsfs}
\usepackage{amsfonts}
\usepackage{amsmath}
\usepackage{array}
\usepackage{verbatim}
\usepackage{epsfig}
\usepackage{graphicx}

\begin{document}
\title{Kinematical correlations for Higgs boson plus High $P_T$ Jet
Production at Hadron Colliders}

\author{Peng Sun}
\affiliation{Nuclear Science Division, Lawrence Berkeley National
Laboratory, Berkeley, CA 94720, USA}
\author{C.-P. Yuan}
\affiliation{Department of Physics and Astronomy, Michigan State University,
East Lansing, MI 48824, USA}
\author{Feng Yuan}
\affiliation{Nuclear Science Division, Lawrence Berkeley National
Laboratory, Berkeley, CA 94720, USA}

\begin{abstract}
We investigate the effect of QCD resummation to kinematical correlations
in the Higgs boson plus high transverse momentum ($P_T$) jet events
produced at hadron colliders. We show that at the complete one-loop order,
the Collins-Soper-Sterman resummation formalism can be applied to derive
the Sudakov form factor, which is found to be independent of jet-finding
algorithm. We compare the singular behavior of resummation calculation
to fixed order prediction in the case that Higgs boson and high $P_T$
jet are produced nearly back-to-back in their transverse momenta,
and find a perfect agreement. The phenomenological importance of the
resummation effect at the LHC is also demonstrated.
\end{abstract}

\maketitle

{\it Introduction.}
Higgs boson discovery at the CERN LHC~\cite{Aad:2012tfa,Chatrchyan:2012ufa} has stimulated new area of
high energy physics research at the colliders, where the precision
Higgs physics is at the frontier. This includes Higgs production and decays
to investigate the coupling between the Higgs boson and all other particles.
The correlation between Higgs and jet production at the LHC will undoubtedly
provide important information on the production and further disentangle
the electroweak coupling of
Higgs boson~\cite{Dawson:1990zj,Djouadi:1991tka,deFlorian:1999zd,Ravindran:2002dc,
Glosser:2002gm,Campbell:2006xx,Campbell:2012am,Boughezal:2013uia,Gangal:2013nxa,Dittmaier:2012vm}.
The goal of this paper is to
build a theoretical framework to reduce the uncertainties in the
Higgs plus jet production at the LHC. The fixed
order perturbative calculations suffer singularities in the back-to-back
correlation region, where the total transverse momentum of
Higgs boson plus jet becomes much smaller than the invariant mass.
Therefore, we have to perform all order soft gluon resummation
to make reliable prediction.

QCD resummation for this process has its own interest in perturbative
QCD. To deal with the divergence in low transverse
momentum hard processes, the so-called transverse momentum, or
Collins-Soper-Sterman (CSS), resummation is employed~\cite{Collins:1984kg}. However, the CSS
resummation has been mainly applied to the color-neutral particle
production, such as inclusive vector boson $W/Z$
and Higgs boson productions. Extension to jet productions in the
final state has been much limited. This is not only because of the technique
issues associated with the jets in the final state, but also because
that the jets carry color and the soft gluon interactions are more
complicated than those for color neutral particle production.
Nevertheless, there have been progresses
in the last few years on the CSS resummation for dijet
production in hadronic collisions~\cite{Banfi:2008qs,Mueller:2013wwa,Sun:2014gfa}.
In this paper, we investigate the CSS resummation for Higgs boson plus one hard jet production,
\begin{equation}
A(P )+B(\bar P)\to H+Jet+X \ ,
\end{equation}
where two incoming hadrons carry momenta $P$ and $\bar P$, respectively.
Because the final state is simpler than that of dijet production, the above
process allows us to study the factorization in great detail.
Extension to W/Z boson plus jet production shall be straightforward, which
are phenomenologically important at the LHC as well.

In the calculations, we apply the effective theory to describe Higgs coupling
to gluons in the large top mass limit:
\begin{equation}
{\cal L}_{eff}=-\frac{\alpha_s}{12 \pi v} F^a_{\mu\nu}F^{a\mu\nu}H,
\label{eq:ggh}
\end{equation}
where $v$ is the vacuum expectation value and $H$ the Higgs field, 
$F^{\mu\nu}$ the gluon field strength tensor and $a$ the color index.
Our final resummation formula can be summarized as
\begin{eqnarray}
\frac{d^4\sigma}
{dy_h dy_j d P_T^2
d^2q_{\perp}}=\sum_{ab}\sigma_0\left[\int\frac{d^2\vec{b}_\perp}{(2\pi)^2}
e^{-i\vec{q}_\perp\cdot
\vec{b}_\perp}W_{ab\to Hc}(x_1,x_2,b_\perp)+Y_{ab\to Hc}\right] \ ,
\end{eqnarray}
where $y_h$ and $y_j$ are rapidities for the Higgs boson and the jet,
$P_T$ for the jet transverse momentum,
 and $\vec{q}_\perp=\vec{P}_{h\perp}+\vec{P}_J$ for the total
transverse momentum of Higgs and the jet.
The first term $W$ contains all order resummation
and the second term $Y$ comes from the fixed order corrections;
$\sigma_0$ represents normalization of the
differential cross section. In this paper, we will take the dominant $gg\to Hg$ channel
as an example to demonstrate how to derive the resummation for $W$ term,
which can be written as
\begin{eqnarray}
W_{gg\to Hg}\left(x_1,x_2,b\right)&=&{H}_{gg\to Hg} (Q)x_1f_g(x_1,\mu=b_0/b_\perp)
x_2f_g(x_2,\mu=b_0/b_\perp) e^{-S_{\rm Sud}(Q^2,b_\perp)} \ ,\label{resum}
\end{eqnarray}
at next-to-leading logarithmic (NLL) level, where $Q^2=s=x_1x_2S$ and represents the hard momentum scale,
$b_0=2e^{-\gamma_E}$, with $\gamma_E$ being the Euler constant.
$f_{a,b}(x,\mu)$ are parton distributions for the incoming
partons $a$ and $b$,
and $x_{1,2}$ are momentum fractions of the incoming hadrons carried by the partons.
Beyond the NLL, a $C$ function associated with the gluon distribution function
will also be included.
The Sudakov form factor can be written as
\begin{eqnarray}
S_{\rm Sud}(Q^2,b_\perp)=\int^{Q^2}_{b_0^2/b_\perp^2}\frac{d\mu^2}{\mu^2}
\left[\ln\left(\frac{Q^2}{\mu^2}\right)A+B+D\ln\frac{1}{R^2}\right]\ , \label{su}
\end{eqnarray}
where $R$ represents the cone sizes for the jet.
Here the parameters $A$, $B$, $D_1$ can be expanded
perturbatively in $\alpha_s$. For $gg\to Hg$ channel,
we have $A=C_A \frac{\alpha_s}{\pi}$,
$B=-2C_A\beta_0\frac{\alpha_s}{\pi}$, and $D=C_A\frac{\alpha_s}{2\pi}$.
The hard coefficient $H$ can be calculated
order by order. From the leading Born diagrams, we have
$H^{(0)}=\left(s^4+t^4+u^4+m_h^8\right)/(stu)$~\cite{Ravindran:2002dc,Glosser:2002gm},
where $s=Q^2$,
$t$ and $u$ are usual Mandelstam variables for the
partonic $2\to 2$ process.

To derive the above resummation, we first calculate $W(b)$ at the
complete one-loop order, and show that it can be
factorized into the parton distributions and soft and hard factors.
The resummation is achieved by solving the associated evolution
equations. The asymptotic behavior at low
imbalance transverse momentum $q_\perp$ is calculated from
the soft and collinear gluon radiations at this order.
This asymptotic result will be checked against the full perturbative
calculations.
Then, we will combine these contribution with those from virtual
graphs and collinear jet contributions to derive the one-loop
result for $W(b)$.

{\it Asymptotic Behavior at Small-$q_\perp$.}
The leading order calculations for the process of Eq.~(1) comes from
the partonic process,
$$g+g\to H+g\ ,$$
which predicts a Delta function at $q_\perp=0$.
At the next-to-leading order (NLO), the real emission diagrams
for $g + g \to H + {\rm jet} + X$  will contribute
to a singular behavior at small-$q_\perp$,
in the associate production of Higgs boson and
high $P_T$ jet, with additional parton radiation.
For the collinear gluon associated with the incoming
gluon distribution, they can be easily evaluated, and they are
proportional to the gluon-to-gluon splitting kernel at one-loop order.
For the soft gluon radiation, we apply the
soft gluon approximation in the limit of $q_\perp\ll Q$, and obtain the following
expression,
\begin{eqnarray}
\int \frac{d^3k_g}{(2\pi)^32E_{k_g}}\delta^{(2)} (q_\perp-k_{g\perp})
\left[\frac{p_1\cdot p_2}{p_1\cdot k_{g}p_2\cdot k_g}
+\frac{k_1\cdot p_2}{k_1\cdot k_{g}p_2\cdot k_g}+\frac{k_1\cdot p_1}{k_1\cdot k_{g}p_1\cdot k_g}\right] \ ,
\end{eqnarray}
where $p_1$, $p_2$ represent the momenta for incoming gluons,
$k_1$ for final state jet, and $k_g$ for the radiated gluon.
However, not all the soft gluon radiation contributes to the finite
$q_\perp$. In particular, if the gluon radiation is within the final state
jet, its contribution has to be excluded.
To evaluate these contributions, we introduce a small
offshellness (which is proportional to the cone size $R$)
for $k_1$ to exclude the gluon radiation inside the jet cone, and further take the
narrow jet approximation (NJA)~\cite{Jager:2004jh}, i.e,
taking the limit of $R\to 0$. In the NJA, this is
equivalent to applying a kinematic cutoff for the radiated gluon.

Adding the soft and collinear gluon radiation together, we derive the
asymptotic behavior at small-$q_\perp$,
\begin{eqnarray}
&&\frac{\alpha_sC_A}{2\pi^2}\frac{1}{q_\perp^2}\int\frac{dx_1'dx_2'}{x_1'x_2'} x_1'g(x_1')x_2'g(x_2')
\left[\left\{\delta(\xi_2-1)\xi_1{\cal P}_{gg}(\xi_1)+(\xi_1\leftrightarrow \xi_2)\right\}\nonumber\right.\\
&&\left.+\delta(\xi_1-1)\delta(\xi_2-1)\left(2\ln\frac{Q^2}{q_\perp^2}-4\beta_0+\ln\frac{1}{R^2}
+\epsilon\left(\frac{1}{2}\ln^2\left(\frac{1}{R^2}\right)+\frac{\pi^2}{6}\right)\right)\right] \ ,\label{soft}
\end{eqnarray}
where $\xi_1=x_1/x_1'$, $\xi_2=x_2/x_2'$, ${\cal P}_{gg}$
is the gluon splitting kernel and $\beta_0=(11-2N_f/3)/12$,
with $N_f$ being the number of effective light quarks.
We have kept the $\epsilon=(4-D)/2$ terms, which
will contribute when Fourier transforming
to $b_\perp$-space.
In Eq.~(\ref{soft}), the first term comes from collinear gluon radiation.
The most important contribution comes from the $\ln({Q^2}/{q_\perp^2})$ term
which is the well-known Sudakov double logarithm in the low
transverse momentum limit. Because of the final state jet
in the process, we also have a jet size dependent term, similar
to dijet production studied in Ref.~\cite{Sun:2014gfa}.

It is important to check the above asymptotic behavior against
the fixed order calculations in the small transverse momentum
limit $q_\perp\ll Q$. In Fig.~1, we plot the comparison between
Eq.~(\ref{soft}) and that from fixed order calculation. We show
the $q_\perp$-dependent differential cross section in the low
transverse momentum region for the typical kinematics at the LHC.
In this plot, we focus on the $gg\to H+Jet$ production channel, and
the jet transverse momentum is in the range between 60 to 100 GeV.
Both the Higgs boson and jet are produced in the central rapidity
region, with $|y|< 0.5$, and the jet
size is set to be $R=0.5$. The full NLO calculation come from the MCFM
code~\cite{Campbell:2010ff}, whereas the asymptotic result from Eq.~(\ref{soft}).
In the numeric calculations, we have adopted the CT10 PDF set~\cite{Gao:2013xoa}.
From this plot, we can clearly see that the asymptotic behavior agrees
well with the fixed order calculation in the low $q_\perp$ region.

\begin{figure}[tbp]
\centering
\includegraphics[width=9cm]{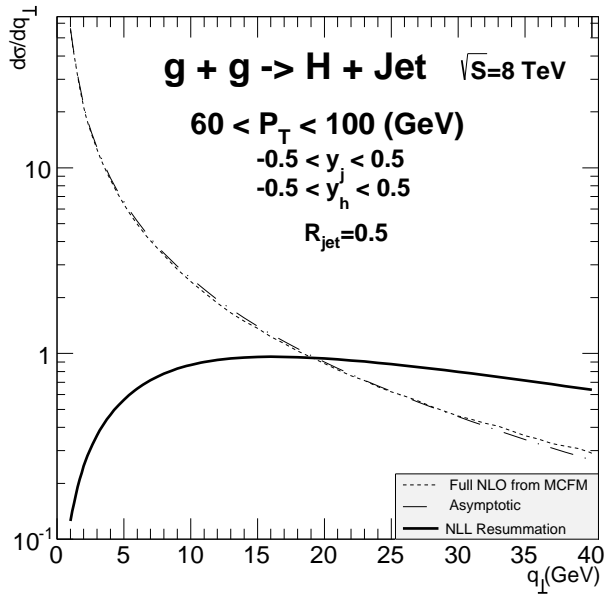}
\caption{Higgs boson plus jet production from $gg\to H+jet$ channel
at the LHC as function of total transverse momentum of the Higgs
boson and the jet. The dotted curve represents the result from the MCFM
code, whereas the dashed curve from asymptotic result of
Eq.~(\ref{soft}). For comparison, the resummed cross section
is shown as the solid curve.}
\label{asymptotic}
\end{figure}

{\it One-loop calculation of $W(b)$.}
To calculate the complete one-loop result for $W(b_\perp)$, we have
to take into account the following three contributions: (a) virtual
graphs contribution to $gg\to Hg$; (b) real gluon contribution associated
to the jet; $(c )$, the collinear and soft gluon radiation contribution to finite $q_\perp$.
Both (a) and (b) contribute to $\delta^{(2)}(q_\perp)$. All these
contributions contain soft divergences, which have to be cancelled out.
In the end, we only have collinear divergences
associated with the incoming two gluons.
To calculate $(c )$, we have to Fourier
transform the $q_\perp$-dependent expression of last section
into $b_\perp$-space.

Calculations of virtual graphs are available in the literature, and
they can be written as~\cite{Ravindran:2002dc,Glosser:2002gm,Schmidt:1997wr}
\begin{eqnarray}
\frac{\alpha_sC_A}{2\pi}\left[-\frac{3}{\epsilon^2}+\frac{1}{\epsilon}\left(2\ln\frac{s}{\mu^2}+\frac{tu}{s\mu^2}\right)
+\cdots\right]
 \ , \label{vir}
\end{eqnarray}
where for simplicity, we only kept the divergent terms.
Collinear gluon associated with the jet is also easy to carry out,
which will depend on the jet algorithm. Following the
anti-$k_t$ algorithm, we obtain the following
contribution for the gluon jet at the one-loop order~\cite{Jager:2004jh,Mukherjee:2012uz}:
\begin{eqnarray}
\frac{\alpha_sC_A}{2\pi}\left[\frac{1}{\epsilon^2}+\frac{1}{\epsilon}\left(2\beta_0-\ln\frac{P_T^2R^2}{\mu^2}
\right)
\!+\frac{1}{2}\ln^2\left(\frac{P_T^2R^2}{\mu^2}\right)\!-2\beta_0\ln\frac{P_T^2R^2}{\mu^2}
\!+\frac{67}{9}\!-\frac{3}{4}\pi^2\!-\frac{23}{54}N_f\right] \ , \label{jet}
\end{eqnarray}
where the divergent and logarithmic terms are independent of jet algorithm, and the
rest of the finite terms depend on the algorithm~\cite{Ji:2004wu}.
We note that the above result includes
contributions from final state gluon splitting into
a gluon pair or a quark-antiquark pair at the NLO, via the
initial state gluon-gluon fusion processes.
At the NLO, we also need to renormalize the
effective $ggH$ coupling, cf. Eq. (\ref{eq:ggh}), which
yields the following contribution:
\begin{equation}
\frac{\alpha_sC_A}{2\pi}\left(\frac{Q^2}{\mu^2}\right)^{-\epsilon}\left(-\frac{3}{\epsilon}\right)2\beta_0 \ .
\end{equation}
Here, we have set the renormalization scale as $Q^2$ to simplify the final expression.

Clearly, the soft divergences ($1/\epsilon^2$ terms) from Eqs.~(\ref{vir}) and (\ref{jet}) are
cancelled out by that in Eq.~(\ref{soft}). The collinear divergence in terms of $(1/\epsilon) \ln (1/R)$
associated with the final state jet is also cancelled out between the jet contributions
from Eqs.~(\ref{soft}) and (\ref{jet}).
In addition, the finite
$\ln^2(P^2_T R^2/\mu^2)$ terms are cancelled out after summing over
Eqs.~(\ref{soft}), (\ref{vir}) and (\ref{jet}).
The above
cancellations provide important cross checks for our derivations. Finally, there are
only divergences coming from the collinear divergences of the gluon distributions.
After renormalizing the gluon distributions from the
incoming hadrons, we obtain the finite contribution at one-loop order,
\begin{eqnarray}
W^{(1)}(b)\!\!&=\!\!&H^{(0)}\frac{\alpha_sC_A}{2\pi}\left\{
\ln\frac{b_0^2}{b^2\bar\mu^2}
\left[\delta(\xi_2-1)\xi_1{\cal P}_{gg}(\xi_1)+(\xi_1\leftrightarrow \xi_2)\right]+\delta(\xi_1-1)\delta(\xi_2-1)\right.\nonumber\\
&\!\!&\left. \!\!\times\!\!\left[-\left(\ln\frac{Q^2b_\perp^2}{b_0^2}\right)^2+\left(4\beta_0-\ln\frac{1}{R^2}\right)\ln\frac{Q^2b_\perp^2}{b_0^2}\right]
\right\}+H^{(1)} \delta(\xi_1-1)\delta(\xi_2-1)\ ,
\end{eqnarray}
where a common integral of the parton distributions as
that in Eq.~(\ref{soft}) is implicit but not shown. In the above result,
the leading and sub-leading logarithmic terms are evident, and the
remaining hard coefficient $H^{(1)}$ is
\begin{eqnarray}
H^{(1)}&=&H^{(0)}\frac{\alpha_sC_A}{2\pi}\left[\ln^2\left(\frac{Q^2}{P_T^2}\right)+2\beta_0\ln\frac{Q^2}{P_T^2R^2}
+\ln\frac{1}{R^2}\ln\frac{Q^2}{P_T^2}-2\ln\frac{-t}{s}\ln\frac{-u}{s}\nonumber\right.\\
&&+\ln^2\left(\frac{\tilde t}{m_h^2}\right)-\ln^2\left(\frac{\tilde t}{-t}\right)+\ln^2\left(\frac{\tilde u}{m_h^2}\right)-\ln^2\left(\frac{\tilde u}{-u}\right)
+2{\rm Li}_2\left(1-\frac{m_h^2}{Q^2}\right)\nonumber\\
&&\left.+2{\rm Li}_2\left(\frac{t}{m_h^2}\right)+2{\rm Li}_2\left(\frac{u}{m_h^2}\right)+\frac{67}{9}+\frac{\pi^2}{2}-\frac{23}{54}N_f\right] +\delta H^{(1)} \ ,
\end{eqnarray}
where $\tilde t=m_h^2-t$, $\tilde u=m_h^2-u$, and $\delta H^{(1)}$ represents terms not proportional
to $H^{(0)}$ and can be
found in Refs.~\cite{Glosser:2002gm,Ravindran:2002dc}. The above will enter
into final resummation result as one-loop correction to Eq.~(\ref{resum}).

{\it TMD Factorization and Resummation.}
In order to carry out the resummation,
we factorize the above one-loop results into
the TMD parton distributions, and soft and hard factors,
following the CSS procedure~\cite{Collins:1984kg}.
In $b_\perp$-space, this factorization can be written as
\begin{eqnarray}
W\left(Q,b_\perp\right)=x_1\,f_g(x_1,b_\perp,\zeta^2,\mu^2,\rho)x_2 f_g(x_2,b_\perp,\bar\zeta^2,\mu^2,\rho)
 {H}_{gg\to Hg}^{TMD}(Q^2,\mu^2,\rho){S}_{gg\to Hg}(b_\perp,\mu^2,\rho) \nonumber \ ,
\end{eqnarray}
where we have followed Ji-Ma-Yuan scheme to define
the TMD gluon distribution $f_g$~\cite{Ji:2004wu}. In this scheme,
an off-light-cone vector $v$ ($\bar v$) is introduced to regulate the
light-cone singularity, $\zeta^2=(2v\cdot P)^2/v^2$ ($\bar \zeta^2=(2\bar v\cdot \bar P)^2/\bar v^2$).
The dependence on $\rho=(2v\cdot \bar v)^2/v^2\bar v^2$ and the factorization scale $\mu$ cancel
out among different factors. An evolution
equation can be derived for the TMD distributions respect to
$\zeta$, from which we will resum large logarithms~\cite{Ji:2004wu}.
The soft factor is defined as
\begin{eqnarray}
S_{gg\to Hg}&=&f_{abc}f_{a'b'c'}\langle 0|
{\cal L}_{vad}^\dagger(b_\perp) {\cal
L}_{\bar vbe} (b_\perp) {\cal L}_{n cf}^\dagger(b_\perp){\cal
L}_{nc'f} (0)
{\cal L}_{\bar vb'e}^\dagger(0) {\cal
L}_{ va'd}(0)  |0\rangle \ ,\label{softf}
\end{eqnarray}
which includes the soft gluon interactions between the final state jet defined
in the $n$-direction (along the jet) and the incoming patrons defined by the two vectors
$v$ and $\bar v$.
A renormalization group equation for the soft factor
can be calculated, and the anomalous dimension is found to be
$\gamma^{(s)}=\frac{\alpha_sC_A}{2\pi}\left(\ln\rho^2+\ln\frac{1}{R^2}\right)$.
The final resummation results of Eq.~(\ref{resum}) were derived by solving
the evolution equations for the parton distributions and the renormalization
group equation for the soft factor.

As an example, in Fig.~1, we show the resummation results, as compared
to the fixed order calculations.
When Fourier transforming the $b_\perp$-expression to obtain the
transverse momentum distribution, we follow the $b_*$ prescription
of CSS resummation~\cite{Collins:1984kg}, and apply the non-perturbative form factors
following the parameterizations in Refs.~\cite{Sun:2012vc}. The final
result is not sensitive to the choice of the non-perturbative form
factor.
In the numeric calculations, we have used parameter $A^{(1,2)}$,
$B^{(1)}$, $D^{(1)}$ and $H^{(0,1)}$ in the resummation formula
Eq.~(\ref{resum}). All these coefficients are obtained from our one-loop
calculation, except that of $A^{(2)}$, for which we argue that the universality
of the gluon distributions will lead to the same $A^{(2)}$ as that for any gluon-gluon
fusion process such as $gg\to H$~\cite{deFlorian:2000pr}. We would like
to emphasize that $H^{(1)}$ correction is of order 1 in the kinematics
shown in Fig.~1, which highlights the importance of next-to-leading
corrections.
From this plot, we can clearly see that the resummation is
important in the kinematic region where the fixed order
calculations have singular behavior. The calculations including
all other channels will be presented in a separate publication.

{\it Conclusions.} In Summary, we have derived all order soft gluon resummation for Higgs
boson plus high energy jet production. The expansion of our
resummation formula agrees well with the fixed order calculations in the
low transverse momentum region of Higgs boson and the jet, where we
showed that resummation effects have to be included to have a reliable
prediction.

Our derivations are based on a complete one-loop perturbative calculation.
The results have been cross checked in various respects.
It demonstrates that the final resummation formalism is consistent in the
framework of CSS resummation.
These results will provide important guidelines for future developments
in electroweak boson plus jet production processes at the LHC.
Extension to Higgs boson (or electroweak boson) plus two jets
production shall be followed as well, which is a potential channel
to investigate the unique production mechanism for Higgs boson
at the collider.

This material is based upon work supported by the U.S. Department of Energy,
Office of Science, Office of Nuclear Physics, under contract number
DE-AC02-05CH11231, and by the U.S. National
Science Foundation under Grant No. PHY-0855561 and PHY-1417326.

\end{document}